\documentclass[11pt,a4paper]{article}

\usepackage{jheppub}

\usepackage{amsmath,amssymb}
\usepackage{enumerate}
\usepackage{amsfonts}

\usepackage{subcaption}
\renewcommand{\figurename}{Fig.}
\usepackage{allpurpose}

\usepackage[marginal,perpage,symbol]{footmisc}
\DefineFNsymbols*{dagfirst}{\dag\ddag*{**}{\dag\dag}{\ddag\ddag}}
\setfnsymbol{dagfirst}

\newcommand{\cP}{\mathcal{P}}
\newcommand{\ahat}{\hat a}

\graphicspath{{./figures}{../figures/}{../selfdual/figures/}}

\everymath{\displaystyle}
\makeatletter
\gdef\@fpheader{}
\makeatother

\usepackage[nottoc,notlof,notlot]{tocbibind}

\usepackage{indentfirst}

\begin{document}

\title{Pseudo-spectral methods in gravitational lensing: Spin precession in loop quantum black hole spacetimes}
\date{\today}

\author[a,*]{Xiankai Pang\note[*]{The corresponding author.}}
\emailAdd{xkpang@cwnu.edu.cn}
\affiliation[a]{\normalfont School of Physics and Astronomy, China West Normal University, Nanchong, 637009, China}

\author[b]{and Xiao-Kan Guo}
\emailAdd{kankuohsiao@whu.edu.cn}
\affiliation[b]{\normalfont School of Mathematics and Physics, Yancheng Institute of Technology, Yancheng 224051, China}

\abstract{
Gravitational lensing is usually treated as an initial value problem: for a given impact parameter of an unbound orbit, one computes the resulting deflection angle. In this paper we advocate the inverse point of view. For a prescribed deflection angle, the orbit is selected by two boundary conditions at spatial infinity, so that the lensing problem becomes a well-defined two-point boundary value problem, with the impact parameter playing the role of a nonlinear eigenvalue. We show that pseudo-spectral methods based on Chebyshev collocation solve this boundary value problem accurately at modest computational cost, and that the parallel transport equations governing the spin precession of the particle reduce to a single linear system on the same grid. As a concrete application, we compute the spin precession of massive particles in backward scattering off two black hole spacetimes from loop quantum gravity, the covariant effective metric and the self-dual loop quantum black hole. We find that the quantum corrections are most pronounced for slowly moving particles, while in the ultra-relativistic limit the universal spin flip is always recovered. The flip itself is universal, but the rate at which it is approached is not. The derivative of $\cos\xi$ with respect to the particle speed at $v=1$ depends on the quantum parameters of both spacetimes. The framework is not restricted to the models studied here and applies to any static, spherically symmetric spacetime.
}


\maketitle

\section{Introduction}
\label{sec:introduction}

Gravitational lensing is nowadays a standard tool in astrophysics and cosmology~\cite{Schneider:1992bmb,Wambsganss:1998gg,Bartelmann:2010fz}, and the deflection angle of an unbound orbit is one of its central quantities. Conventionally it is computed as an initial value problem: for a fixed impact parameter $b$ one evaluates the azimuthal angle swept along the orbit, either by integrating the geodesic equations numerically or by expanding the deflection integral in the weak field or the strong deflection limit~\cite{Bozza:2002zj,Bozza:2010xqn,Virbhadra:1999nm,Frittelli:1999yf,Claudel:2000yi,Eiroa:2002mk,Bozza:2006nm}.

In this paper we advocate the inverse point of view. For a prescribed deflection angle, the orbit is selected by two boundary conditions at spatial infinity, so that lensing becomes a well-defined two-point boundary value problem with the impact parameter itself appearing as a nonlinear eigenvalue. The formulation is particularly natural for backward scattering, where the deflection angle takes the values $(2n-1)\pi$, $n=1,2,3,\dots$, and the problem is posed on the fixed interval $[0,2n\pi]$. It is also ideally suited to pseudo-spectral methods, which approximate the solution on a small non-uniform grid and turn the differential equation into a finite algebraic system with exponential (spectral) convergence~\cite{Boyd:1989cfs,Grandclement:2007sb}. A standard tool in numerical relativity, holography and quasinormal-mode computations~\cite{Guo:2016hiw,Andrade:2017jmt,Jansen:2017oag,Gourgoulhon:2001ec,Dias:2011at,Berti:2009kk,Konoplya:2011qq}, the pseudo-spectral method has rarely been applied to gravitational lensing; one purpose of this paper is to show, in a concrete physical setting, how well it performs there.

The physical problem we choose is the precession of the particle spin in backward scattering. Backward (glory) scattering is special for particles carrying spin. Partial wave analyses show a diverging glory spot for scalar waves, but none for massless particles with spin~\cite{Matzner:1985rjn,Dolan:2018nzc,Glampedakis:2001cx,Crispino:2009ki,Dolan:2007ut}. The absence has a classical explanation~\cite{Pang:2024tco}, as backward scattering always reverses the spin of a massless particle, so that contributions from orbits in perpendicular planes cancel exactly. For massive particles the cancellation is incomplete, and the precession angle depends on the spacetime geometry and the particle speed $v$, especially for $v\ll1$~\cite{Geng:2026xcs}; the effect is also a classic probe of relativistic gravity~\cite{Schiff:1960gi,Schiff:1960gh,Everitt:2011hp,Bini:2005nt}. Its computation requires solving the orbit first and then two coupled linear transport equations along it, a structure ideally suited to the boundary value formulation and the pseudo-spectral method.

We apply the framework to two black hole spacetimes obtained in loop quantum gravity (LQG): the covariant effective metric of~\cite{Zhang:2024khj} (the solution denoted BH-II there, with the underlying covariant dynamics further developed in~\cite{Lewandowski:2022zce,Livine:2025soz}) and the self-dual black hole of~\cite{Modesto:2009ve}, obtained by polymerizing the Schwarzschild interior. Both belong to a long line of LQG-inspired metrics~\cite{Gambini:2013ooa,Modesto:2004xx,Modesto:2005zm,Modesto:2008im,Ashtekar:2018lag,Ashtekar:2018cay,Olmedo:2017lvt,Perez:2017cmj,Rovelli:2014ssa}. They have been constrained by Solar System observations~\cite{Huang:2025ggs}, and studied through their photon spheres and shadows~\cite{Han:2026bgn} (properties probed by the horizon-scale images of the Event Horizon Telescope~\cite{EventHorizonTelescope:2019dse,EventHorizonTelescope:2022wkp,Cardoso:2019rvt}), as well as through geodesic motion and waveforms~\cite{Chen:2025baz,Chen:2025aqh,Li:2024afr,Li:2025zdt,Wang:2024iwt}.

In this paper, we compute the cosine of the spin precession angle~\cite{Pang:2024tco,Geng:2026xcs}, $\cos\xi$, for massive particles backward scattered by these two spacetimes, as a function of the quantum parameters and the particle speed $v$, including the slope $\dd(\cos\xi)/\dd v$ at $v=1$. For generality, all the equations below are written for a general static, spherically symmetric line element with three free functions $f$, $g$ and $h$, and each spacetime enters only through the specification of these functions. We find that the quantum corrections are most pronounced for slowly moving particles and disappear as $v\to1$, where the universal spin flip $\cos\xi=-1$ is recovered~\cite{Pang:2024tco,Geng:2026xcs}. The approach to the flip is not universal. In fact, the slope $\dd(\cos\xi)/\dd v$ at $v=1$ varies visibly with the quantum parameters of both spacetimes. The spin precession of slow massive particles thus carries clean imprints of the underlying quantum geometry.

The paper is organized as follows. Sec.~\ref{sec:bvp} formulates backward lensing as a boundary value problem and derives the orbit and spin precession equations; Sec.~\ref{sec:imppseudo} describes the pseudo-spectral implementation and its validation; Sec.~\ref{sec:spinprecession} presents the numerical results; and Sec.~\ref{sec:discussion} concludes. A brief review of pseudo-spectral methods and Chebyshev polynomials is collected in appendix~\ref{sec:pseudospectral}. We use units $G=c=1$ and the signature $(+,-,-,-)$.

\section{Backward lensing as a boundary value problem}
\label{sec:bvp}

\subsection{The loop quantum black hole metrics}
\label{sec:metrics}

We consider static, spherically symmetric spacetimes with the line element
\begin{IEEEeqnarray}{rCl}
	\dd s^2 &=& f(r)\dd t^2-g(r)\dd r^2 - h(r)(\dd\theta^2+\sin^2\theta\dd\varphi^2) \quad , \label{eq:metric}
\end{IEEEeqnarray} 
and restrict ourselves to asymptotically flat spacetimes, i.e.\ $f,g\to1$ and $h/r^2\to1$ as $r\to\infty$. Both loop quantum black hole spacetimes studied in this paper are of this form.

The covariant effective metric of~\cite{Zhang:2024khj}, the second of the two solutions constructed there (denoted BH-II in that reference), corresponds to
\begin{IEEEeqnarray}{rCl}
	f(r) &=& 1- \frac{2M}{r} \quad , \qquad g(r)=\frac{1}{f(r)\mu(r)} \quad , \qquad h(r)=r^2 \quad , \nonumber\\
	\mu(r) &=& 1+\frac{\zeta^2}{r^2}\left(1-\frac{2M}{r}\right) \quad . \label{eq:metricBHII}
\end{IEEEeqnarray} 
Here $M$ is the black hole mass and $\zeta$ is the quantum parameter, proportional to the Planck length; the metric reduces to Schwarzschild as $\zeta\to0$~\cite{Zhang:2024khj}, and $\mu>0$ outside the horizon at $r=2M$. Note that $fg=1/\mu\neq1$, i.e.\ $g_{tt}\neq-1/g_{rr}$, while the areal radius still coincides with the radial coordinate.

The self-dual loop quantum black hole of~\cite{Modesto:2009ve} corresponds to
\begin{IEEEeqnarray}{rCl}
	f(r) &=& \frac{(r-r_+)(r-r_-)(r+r_\star)^2}{r^4+a_0^2} \quad , \nonumber\\
	g(r) &=& \frac{(r+r_\star)^2\left(r^4+a_0^2\right)}{(r-r_+)(r-r_-)\,r^4} \quad , \qquad
	h(r) \;=\; r^2+\frac{a_0^2}{r^2} \quad , \label{eq:metricSD}
\end{IEEEeqnarray} 
with
\begin{IEEEeqnarray}{rCl}
	r_+ &=& 2m \quad , \qquad r_-=2m\cP^2 \quad , \qquad r_\star=2m\cP \quad , \qquad a_0=\frac{A_{\rm Min}}{8\pi} \quad , \label{eq:sdparams}
\end{IEEEeqnarray} 
where $A_{\rm Min}$ is the minimum area gap of LQG~\cite{Ashtekar:1997yu} and 
\begin{IEEEeqnarray}{rCl}
	\cP(\delta_b) &=& \frac{\sqrt{1+\gamma^2\delta_b^2}-1}{\sqrt{1+\gamma^2\delta_b^2}+1}\in[0,1) \label{eq:polP}
\end{IEEEeqnarray} 
is a function of the polymeric parameter $\delta_b$ ($\gamma$ is the Barbero-Immirzi parameter). Two features distinguish this spacetime from the covariant metric. First,
\begin{IEEEeqnarray}{rCl}
	f(r)g(r) &=& \frac{(r-r_+)^2(r-r_-)^2r^4}{\left(r^4+a_0^2\right)^2}\neq1 \quad , \label{eq:fgSD}
\end{IEEEeqnarray} 
so $fg\neq1$ here as well. Second, the areal radius $\sqrt{h}=r\sqrt{1+a_0^2/r^4}$ differs from the radial coordinate and has a minimum $\sqrt{h}=\sqrt{2a_0}$ at $r=\sqrt{a_0}$, the wormhole throat of the self-dual solution, which is invariant under the self-duality map $r\to a_0/r$~\cite{Modesto:2009ve}. The asymptotic expansion $f=1-2m(1-\cP)^2/r+\mathcal O(r^{-2})$ shows that the ADM mass is
\begin{IEEEeqnarray}{rCl}
	M &=& m\left(1-\cP\right)^2 \quad . \label{eq:admSD}
\end{IEEEeqnarray} 
Schwarzschild spacetime is recovered only when both $a_0\to0$ and $\cP\to0$. The interior structure of these spacetimes, which is not relevant for the scattering problem studied here, is analysed in detail in~\cite{Zhang:2024khj,Modesto:2009ve}.

In the following we use the dimensionless inverse radius
\begin{IEEEeqnarray}{rCl}
	u &=& \frac{M}{r} \quad , \label{eq:udef}
\end{IEEEeqnarray} 
so that $u=0$ is spatial infinity, and we measure all quantities in units of the ADM mass $M$, i.e.\ we set $M=1$. The metric functions written as functions of $u$ are understood in these units, e.g.\ $h(u)\equiv h(M/u)/M^2$. The remaining dimensionless quantum parameters are
\begin{IEEEeqnarray}{rCl}
	\alpha &=& \frac{\zeta}{M} \qquad \text{(covariant)} \quad , \qquad \ahat=\frac{a_0}{M^2}\quad\text{and}\quad \cP \qquad \text{(self-dual)} \quad . \label{eq:ualpha}
\end{IEEEeqnarray} 
Explicitly, the metric functions of the two spacetimes read
\begin{IEEEeqnarray}{rCl}
	\text{covariant}: \quad f(u) &=& 1-2u \quad , \qquad g(u)=\frac{1}{\left(1-2u\right)\mu(u)} \quad , \nonumber\\
	\mu(u) &=& 1+\alpha^2u^2\left(1-2u\right) \quad , \qquad h(u)=\frac{1}{u^2} \quad , \label{eq:fghBHII}
\end{IEEEeqnarray} 
and
\begin{IEEEeqnarray}{rCl}
	\text{self-dual}: \quad f(u) &=& \frac{\left(1-r_+u\right)\left(1-r_-u\right)\left(1+r_\star u\right)^2}{1+\ahat^2u^4} \quad , \nonumber\\
	g(u) &=& \frac{\left(1+r_\star u\right)^2\left(1+\ahat^2u^4\right)}{\left(1-r_+u\right)\left(1-r_-u\right)} \quad , \nonumber\\
	h(u) &=& \frac{1+\ahat^2u^4}{u^2} \quad , \label{eq:fghSD}
\end{IEEEeqnarray} 
where the characteristic radii in units of the ADM mass are
\begin{IEEEeqnarray}{rCl}
	r_+ &=& \frac{2}{\left(1-\cP\right)^2} \quad , \qquad r_-=\frac{2\cP^2}{\left(1-\cP\right)^2} \quad , \qquad r_\star=\frac{2\cP}{\left(1-\cP\right)^2} \quad . \label{eq:sdradii}
\end{IEEEeqnarray} 
Note that at fixed ADM mass the outer horizon moves outward as $\cP$ grows, so increasing $\cP$ pushes the whole strong-field region to larger radii.

\subsection{The orbit equation and the boundary value problem}
\label{sec:orbit}

We consider a massive particle moving on a timelike geodesic in the equatorial plane $\theta=\pi/2$. The conserved energy and angular momentum per unit mass are
\begin{IEEEeqnarray}{rCl}
	E_0 &=& \frac{1}{\sqrt{1-v^2}} \quad , \qquad L=E_0\,\eta M \quad , \label{eq:E0L}
\end{IEEEeqnarray} 
where $v$ is the particle speed at infinity and $\eta=bv/M$ with $b$ the impact parameter~\cite{Pang:2018jpm,Pang:2024tco}. For the general metric \eqref{eq:metric} the geodesic equation for $u(\varphi)$ reduces to the first integral
\begin{IEEEeqnarray}{rCl}
	\left(\frac{\dd u}{\dd\varphi}\right)^2 &=& \frac{u^4h(u)^2}{\eta^2 g(u)}\left[\frac{1}{f(u)}-\left(1-v^2\right)-\frac{\eta^2}{h(u)}\right] \quad , \label{eq:firstintegral}
\end{IEEEeqnarray} 
and, differentiating once, to the second order orbit equation
\begin{IEEEeqnarray}{rCl}
	\eta^2\frac{\dd^2u}{\dd\varphi^2} &=& \frac{1}{2}\frac{\dd}{\dd u}\left\{\frac{u^4h(u)^2}{g(u)}\left[\frac{1}{f(u)}-\left(1-v^2\right)-\frac{\eta^2}{h(u)}\right]\right\} \quad . \label{eq:orbit}
\end{IEEEeqnarray} 
For Schwarzschild spacetime ($f=1/g=1-2u$, $h=1/u^2$) this reproduces the familiar Binet-type equation $u''+u=(1-v^2)/\eta^2+3u^2$. Explicitly, for the two spacetimes of this paper Eq.~\eqref{eq:firstintegral} reads
\begin{IEEEeqnarray}{rCl}
	\text{covariant}: \quad \left(\frac{\dd u}{\dd\varphi}\right)^2 &=& \frac{\mu(u)}{\eta^2}\left[1-f(u)\left(1-v^2+\eta^2u^2\right)\right] \quad , \label{eq:firstintBHII}
\end{IEEEeqnarray} 
and
\begin{IEEEeqnarray}{rCl}
	\text{self-dual}: \quad \left(\frac{\dd u}{\dd\varphi}\right)^2 &=& \frac{\left(1+\ahat^2u^4\right)^2}{\eta^2\left(1+r_\star u\right)^4}-\frac{\left(1-v^2\right)\left(1+\ahat^2u^4\right)\left(1-r_+u\right)\left(1-r_-u\right)}{\eta^2\left(1+r_\star u\right)^2} \nonumber\\
	&& \quad -\frac{u^2\left(1-r_+u\right)\left(1-r_-u\right)}{\left(1+r_\star u\right)^2} \quad , \label{eq:firstintSD}
\end{IEEEeqnarray} 
and in both cases the right-hand side of the orbit Eq.~\eqref{eq:orbit} is a rational function of $u$. Note that for massless particles ($v=1$) the parameter $\eta$ drops out of \eqref{eq:orbit} only if $u^4h^2/(gf)$ is a constant, as happens when $fg=1$ and $h=r^2$ (Schwarzschild); for the covariant metric this combination equals $\mu(u)$ and for the self-dual black hole $\left(1+\ahat^2u^4\right)^2/\left(1+r_\star u\right)^4$, so $\eta$ enters the orbit equation even at $v=1$ for both spacetimes.

A scattering orbit starts and ends at spatial infinity. If the particle sweeps a total azimuthal angle $2\varphi_0$ before returning to infinity, the orbit satisfies
\begin{IEEEeqnarray}{rCl}
	u(0) &=& 0 \quad , \qquad u(2\varphi_0)=0 \quad , \label{eq:orbitbc}
\end{IEEEeqnarray} 
and the deflection angle is $\Delta\varphi=2\varphi_0-\pi$~\cite{Pang:2018jpm,Pang:2024tco}. Backward scattering corresponds to $\Delta\varphi=(2n-1)\pi$ with $n=1,2,3,\dots$, i.e.\ the particle returns to infinity after sweeping $2\varphi_0=2n\pi$, possibly winding around the black hole several times. In this case the conditions \eqref{eq:orbitbc} are imposed on the fixed interval $[0,2n\pi]$, and the orbit Eq.~\eqref{eq:orbit} becomes a two-point boundary value problem. Since $\eta$ is not known a priori, one more condition is needed. It is supplied by the first integral \eqref{eq:firstintegral} evaluated at the boundary. Using the asymptotic conditions $f,g\to1$ and $u^2h\to1$ as $u\to0$, valid for both spacetimes, one finds
\begin{IEEEeqnarray}{rCl}
	\eta^2\left[u'(0)\right]^2 &=& v^2 \quad . \label{eq:etacondition}
\end{IEEEeqnarray} 
Equations \eqref{eq:orbit}, \eqref{eq:orbitbc} and \eqref{eq:etacondition} form a closed system, with the standard counting for a two-point boundary value problem with one free parameter.
Together they determine the orbit $u(\varphi)$ and the normalized impact parameter $\eta$ simultaneously, with $\eta^2$ playing the role of a nonlinear eigenvalue. This is complementary to the conventional initial value formulation, where $\eta$ is prescribed and the deflection angle is computed. %

\subsection{Spin precession along the orbit}
\label{sec:spineq}

Let the spin of the particle be described by a vector $S^\mu$ which is parallel transported along the geodesic and is orthogonal to the four-velocity, $S^\mu U_\mu=0$~\cite{Pang:2024tco,Schiff:1960gi,Weinberg:1972kfs}. For motion in the equatorial plane, the $S^\theta$ component decouples and can be set to zero, $S^t$ is eliminated by the orthogonality relation, and the remaining two components are conveniently combined into the rescaled variables
\begin{IEEEeqnarray}{rCl}
	\tilde S^r &\equiv& \sqrt{f(r)g(r)}\,S^r \quad , \qquad \bar S^\varphi\equiv \sqrt{h(r)}\,S^\varphi \quad , \label{eq:spinrescale}
\end{IEEEeqnarray} 
with which the terms proportional to the radial velocity in the transport equations cancel identically. For the general metric \eqref{eq:metric} we find the two coupled first order equations
\begin{IEEEeqnarray}{rCl}
	\frac{\dd \tilde S^r}{\dd\varphi} &=& K_1\,\bar S^\varphi \quad , \qquad \frac{\dd \bar S^\varphi}{\dd\varphi}=-K_2\,\tilde S^r \quad , \label{eq:spingen}
\end{IEEEeqnarray} 
with
\begin{IEEEeqnarray}{rCl}
	K_1 &=& \frac{1}{2}\sqrt{\frac{f(r)h(r)}{g(r)}}\left(\frac{h'(r)}{h(r)}-\frac{f'(r)}{f(r)}\right) \quad , \qquad K_2=\frac{h'(r)}{2\sqrt{f(r)g(r)h(r)}} \quad , \label{eq:K1K2}
\end{IEEEeqnarray} 
where the prime denotes $\dd/\dd r$ and the coefficients are evaluated on the orbit $r=M/u(\varphi)$. Since $\sqrt{fg}\to1$ and $\sqrt{h}/r\to1$ at spatial infinity for both spacetimes, $(\tilde S^r,\bar S^\varphi)$ coincide with the physical components $(S^r,rS^\varphi)$ at the two ends of the orbit, so the initial conditions and the definition \eqref{eq:cosxi} of $\xi$ below are unaffected by the rescaling. For Schwarzschild spacetime the coefficients reduce to $K_1=1-3u$, $K_2=1$, i.e.\
\begin{IEEEeqnarray}{rCl}
	\frac{\dd S^r}{\dd\varphi} &=& (1-3u)\,\bar S^\varphi \quad , \qquad \frac{\dd \bar S^\varphi}{\dd\varphi}=-S^r \quad , \label{eq:spinSch}
\end{IEEEeqnarray} 
so that in the flat limit $u\to0$ the spin simply rotates together with the orbiting frame spanned by $(\bm{e}_r,\bm{e}_\varphi)$. For the covariant metric one has $\sqrt{fg}=1/\sqrt{\mu}$, and
\begin{IEEEeqnarray}{rCl}
	\text{covariant}: \quad K_1 &=& (1-3u)\sqrt{\mu(u)} \quad , \qquad K_2=\sqrt{\mu(u)} \quad , \label{eq:spinBHII}
\end{IEEEeqnarray} 
while for the self-dual black hole
\begin{IEEEeqnarray}{rCl}
	\text{self-dual}: \quad K_1 &=& \frac{u\left(1-r_+u\right)\left(1-r_-u\right)}{2\sqrt{1+\ahat^2u^4}}\left[\frac{2}{u}-\frac{r_+}{1-r_+u}-\frac{r_-}{1-r_-u}\right. \nonumber\\
	&& \qquad\qquad \left. +\,\frac{2r_\star}{1+r_\star u}-\frac{8\ahat^2u^3}{1+\ahat^2u^4}\right] \quad , \nonumber\\
	K_2 &=& \frac{1-\ahat^2u^4}{\left(1+r_\star u\right)^2\sqrt{1+\ahat^2u^4}} \quad , \label{eq:spinSD}
\end{IEEEeqnarray} 
both of which reduce to the Schwarzschild coefficients as $(\ahat,\cP)\to(0,0)$. Already at this level one sees a structural difference between the two spacetimes. For the covariant metric the quantum corrections enter the transport coefficients only at order $\alpha^2u^2$, through $\sqrt{\mu}=1+\frac12\alpha^2u^2+\dots$, whereas for the self-dual black hole the polymeric parameter enters already at linear order in $u$,
\begin{IEEEeqnarray}{rCl}
	K_1 &=& 1-\left(3+4\cP+\mathcal O(\cP^2)\right)u+\mathcal O(u^2) \quad , \qquad K_2=1-2r_\star u+\mathcal O(u^2) \quad , \label{eq:Ksmallu}
\end{IEEEeqnarray} 
while the minimum-area parameter appears only at order $\ahat^2u^4$. Larger deviations of the spin precession are therefore to be expected for the self-dual black hole at moderate $\cP$ than for the covariant metric at comparable $\alpha$.

We take the initial spin at infinity to be aligned with $\bm{e}_\varphi$, i.e.\ $\tilde S^r(0)=0$ and $\bar S^\varphi(0)=E_0$ (the overall normalization is irrelevant since the equations are linear and only the direction matters). At the final point, which is again at infinity, the spin measured in the rest frame of the particle is obtained from $(\tilde S^r,\bar S^\varphi)$ by the radial boost with factor $E_0$, and the precession angle $\xi$, defined as the angle between the initial and the final spin in the rest frame, satisfies~\cite{Pang:2024tco}
\begin{IEEEeqnarray}{rCl}
	\cos\xi &=& \frac{E_0\,\bar S^\varphi(2n\pi)}{\sqrt{\left[\tilde S^r(2n\pi)\right]^2+E_0^2\left[\bar S^\varphi(2n\pi)\right]^2}} \quad . \label{eq:cosxi}
\end{IEEEeqnarray} 
As we will see in Sec.~\ref{sec:spinprecession}, the numerical results suggest that, even in the presence of quantum effects, one always finds $\cos\xi=-1$ for massless particles after backward scattering. The spin is reversed, leading to the absence of the glory spot~\cite{Crispino:2009xt,Pang:2024tco,Dolan:2018nzc}.

\section{The pseudo-spectral implementation}
\label{sec:imppseudo}

We now describe how the boundary value problem of Sec.~\ref{sec:bvp} is solved with the pseudo-spectral method. We discuss in turn the discretization of the orbit equation, the solution of the spin transport equations, and the validation of the solutions. The method itself is reviewed in appendix~\ref{sec:pseudospectral}, and a fully explicit worked example, in which the differentiation matrix is small enough to be written down, is given in appendix~\ref{sec:workedexample}.

\paragraph{The orbit.} The interval $[0,2n\pi]$ is mapped to the standard spectral interval $x\in[-1,1]$ by $\varphi=n\pi(1-x)$ and covered with $N+1$ Chebyshev-Gauss-Lobatto collocation points $x_j=\cos(j\pi/N)$, $j=0,\dots,N$~\cite{Boyd:1989cfs}. A function is represented by its values on the grid, and its derivatives by multiplication of the vector of values with the differentiation matrices $D^{(1)}$ and $D^{(2)}$ (see appendix~\ref{sec:pseudospectral}). Using the reflection symmetry $u(\varphi)=u(2n\pi-\varphi)$ of the orbit about the turning point, only the first half of the grid is independent. The full vector of values is reconstructed by even reflection, and the boundary condition $u(2n\pi)=0$ is automatic. The independent unknowns are therefore the $(N+1)/2$ values $u_j=u(\varphi_j)$ on the half grid, together with the squared normalized impact parameter $\eta^2$. They are matched by an equal number of algebraic equations, namely the collocated orbit Eq.~\eqref{eq:orbit} on the half grid with its first row replaced by the boundary condition $u_0=0$, and the supplementary condition $\eta^2\left[\sum_jD^{(1)}_{0j}u_j\right]^2=v^2$ obtained from the first integral \eqref{eq:etacondition}. In this way the continuum counting of conditions of Sec.~\ref{sec:orbit}, two boundary conditions for the second order orbit equation plus one condition for the free parameter, is reproduced exactly at the discrete level. The algebraic system is square, and the well-posedness of the boundary value problem translates into the invertibility of its Jacobian, which we monitor throughout. The system is solved by Newton's method with the analytically computed Jacobian, and a solution is accepted only when the residual norm of the full system is below $10^{-8}$ (working precision $25$ digits). We take $N=73$ for the $\Delta\varphi=\pi$ channel and $N=53$ for the higher channels in all scans at $v=10^{-3}$ and in the velocity scan for the self-dual black hole, and $N=101$ in the velocity scan for the covariant metric at $\alpha=2$ so that $\eta^2$ remains converged also for the near-capture orbits at high $v$. We set $M=1$ in all numerical computations. The initial guess for $u$ is a smooth bump profile and for $\eta^2$ the value of the corresponding Schwarzschild orbit. When Newton's method does not converge directly, the target point is reached by continuation, chaining the converged solution at a nearby value of the parameters as the next initial guess and bisecting the parameter gap whenever a step fails.

\paragraph{The spin.} Once $u(\varphi)$ and $\eta$ are known, the transport equations \eqref{eq:spingen} with \eqref{eq:spinBHII} or \eqref{eq:spinSD} are linear in $(\tilde S^r,\bar S^\varphi)$, with coefficients $K_{1,2}$ that are simply evaluated at the grid values of the orbit. Collocating them on the same grid therefore produces a single linear system for the spin values at the grid points; the first row of each equation block is replaced by the initial conditions $\tilde S^r(0)=0$, $\bar S^\varphi(0)=E_0$, the same row-replacement prescription used for the boundary conditions of the orbit equation. One linear solve then gives the spin along the whole orbit, and the precession angle follows from \eqref{eq:cosxi}. No Newton iteration is needed for this step, and its cost is negligible compared with the orbit solve.

Two degeneracies of the boundary value problem deserve mention. First, for the $\Delta\varphi=\pi$ channel at very small $v$ the problem becomes nearly singular (the return angle flattens as a function of $\eta^2$), which is handled by the continuation in $v$ described above together with an increased number of collocation points. Second, near this fold the orbit can touch $u=0$ at $\varphi=2n\pi$ without crossing it, and the discrete system then admits solutions whose continuum return angle differs from $2n\pi$ (``phantom'' solutions). Every point shown in the figures has therefore been verified against the continuum orbit equation. The return angle is computed by direct quadrature of the first integral \eqref{eq:firstintegral},
\begin{IEEEeqnarray}{rCl}
	\Phi &=& 2\int_0^{u_{\rm t}}\frac{\dd u}{\sqrt{\left[u'(u)\right]^2}} \quad , \label{eq:returnangle}
\end{IEEEeqnarray} 
with the integrand given by the right-hand side of \eqref{eq:firstintegral} and $u_{\rm t}$ the turning point, i.e.\ the smallest positive root of the right-hand side below the horizon; a point is accepted only if $\Phi$ agrees with $2n\pi$ and the resulting $\cos\xi$ agrees with the shooting integration of the transport equations. The Schwarzschild-connected branch is followed by predictive tracking on the return-angle function, selecting the root nearest to a log-linear prediction and refining it by bisection. This is essential for the self-dual black hole, where the root moves fast with $\cP$ at fixed ADM mass, so that plain parameter chaining would jump between branches.

\section{Spin precession in the loop quantum black hole spacetimes}
\label{sec:spinprecession}

\begin{figure}[htpb]
	\centering
	\includegraphics[width=0.62\textwidth]{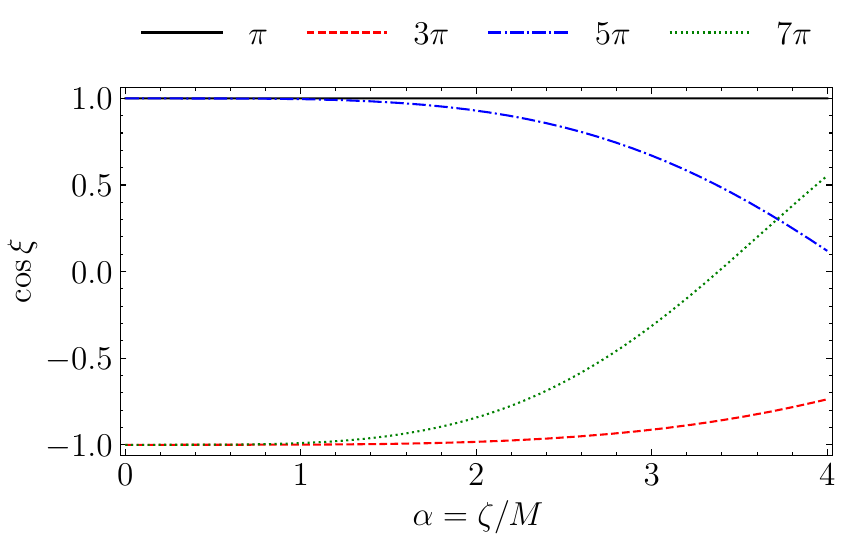}
	\caption{$\cos\xi$  as a function of the quantum parameter $\alpha=\zeta/M$ of the covariant metric at $v=10^{-3}$, for the deflection angles $\Delta\varphi=\pi$ (black solid), $3\pi$ (red dashed), $5\pi$ (blue dash-dotted) and $7\pi$ (green dotted). As $\alpha$ is switched off the Schwarzschild values $\cos\xi=(-1)^{n+1}$ are recovered.}
	\label{fig:cosxiq}
\end{figure}

\begin{figure}[htpb]
	\centering
	\begin{subfigure}{0.48\textwidth}
		\centering
		\includegraphics[width=1.0\textwidth]{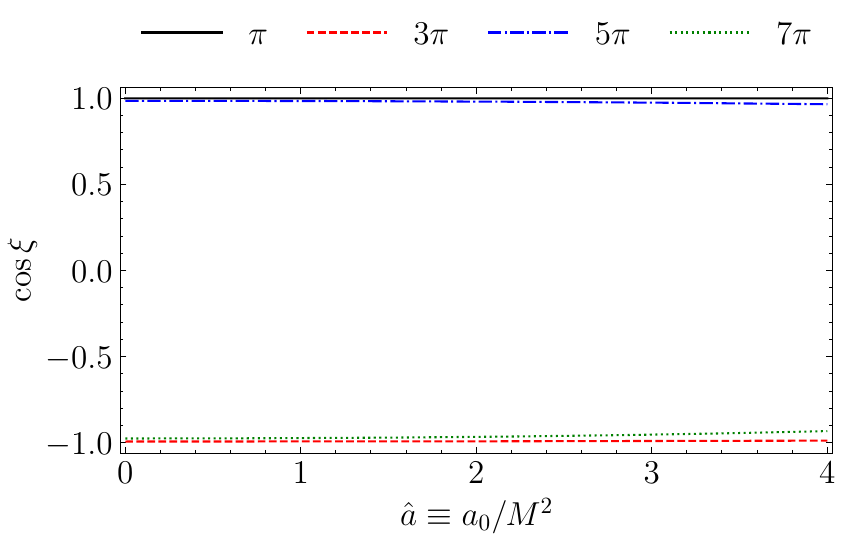}
		\caption{fixed $\cP=1/4$}
		\label{fig:cosxiqSDa}
	\end{subfigure}\hfill
	\begin{subfigure}{0.48\textwidth}
		\centering
		\includegraphics[width=1.0\textwidth]{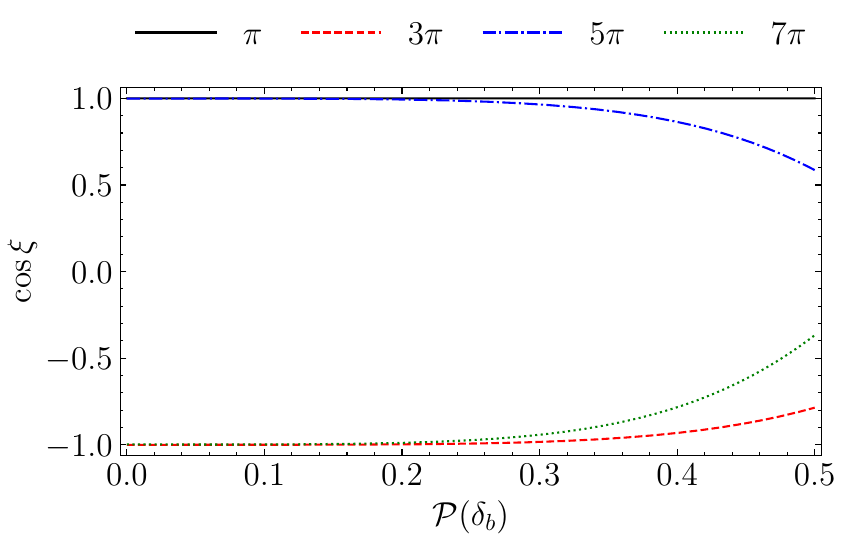}
		\caption{fixed $\ahat=1$}
		\label{fig:cosxiqSDb}
	\end{subfigure}
	\caption{$\cos\xi$ of the self-dual black hole at $v=10^{-3}$, for the same deflection angles and the same line styles as in Fig.~\ref{fig:cosxiq}. Fig.~\ref{fig:cosxiqSDa}: the dependence on the minimum-area parameter $\ahat=a_0/M^2$ at fixed $\cP=1/4$; Fig.~\ref{fig:cosxiqSDb}: the dependence on the polymeric parameter $\cP$ at fixed $\ahat=1$. As the quantum parameters are switched off the Schwarzschild values $\cos\xi=(-1)^{n+1}$ are recovered.}
	\label{fig:cosxiqSD}
\end{figure}

We present the results for the four backward channels $\Delta\varphi=\pi$, $3\pi$, $5\pi$ and $7\pi$ ($n=1,2,3,4$). Figs.~\ref{fig:cosxiq} and~\ref{fig:cosxiqSD} show $\cos\xi$ as a function of the quantum parameters at the small speed $v=10^{-3}$, namely as a function of $\alpha=\zeta/M$ for the covariant metric, and of $\ahat=a_0/M^2$ (at fixed $\cP=1/4$) and $\cP$ (at fixed $\ahat=1$) for the self-dual black hole. Fig.~\ref{fig:cosxiv} shows $\cos\xi$ as a function of $v$, at fixed $\alpha=2$ for the covariant metric and at fixed $(\ahat,\cP)=(1,1/4)$ for the self-dual black hole.

With the exception of the $\pi$ channel, all these orbits lie in the strong-deflection regime, as table~\ref{tab:periapsis} shows. At $v=10^{-3}$ the periapsis for the $3\pi$, $5\pi$ and $7\pi$ channels lies just outside the marginally bound circular orbit, both for the covariant metric, where it keeps the Schwarzschild value $r_c=4M$, and for the self-dual black hole, where it is shifted outward by the quantum deformation. The $\pi$ channel is the marginal case and is a near-parabolic orbit with a periapsis far outside the strong-field region. As $v$ increases the periapsis moves closer to the photon sphere ($r_{\rm ph}=3M$ for the covariant metric), and the impact parameter approaches the critical value for capture.

\begin{table}[htpb]
	\centering
	\begin{tabular}{c|c|c}
		\hline
		$\Delta\varphi$ & covariant, $\alpha=2$ & self-dual, $(\ahat,\cP)=(1,1/4)$ \\
		\hline
		$\pi$ & $224\,M$ & $406.8\,M$ \\
		$3\pi$ & $4.36\,M$ & $8.72\,M$ \\
		$5\pi$ & $4.03\,M$ & $7.64\,M$ \\
		$7\pi$ & $4.00\,M$ & $7.50\,M$ \\
		\hline
	\end{tabular}
	\caption{The periapsis $r_{\rm min}$ of the orbits at $v=10^{-3}$ for the four backward channels. The $3\pi$, $5\pi$ and $7\pi$ orbits pass just outside the marginally bound circular orbit for slow massive particles and are therefore strong-field, near-capture orbits. The $\pi$ channel is a near-parabolic orbit far from the black hole.}
	\label{tab:periapsis}
\end{table}

\subsection{Dependence on the quantum parameters}
\label{sec:alphadep}

At $v=10^{-3}$ the particle is deep in the non-relativistic regime. As the quantum parameters are switched off ($\alpha\to0$ for the covariant metric, $(\ahat,\cP)\to(0,0)$ for the self-dual black hole), all curves approach the Schwarzschild non-relativistic values $\cos\xi=(-1)^{n+1}$, namely $+1$ for $\Delta\varphi=\pi$, $5\pi$ and $-1$ for $\Delta\varphi=3\pi$, $7\pi$~\cite{Pang:2024tco}.

For the covariant metric (Fig.~\ref{fig:cosxiq}), the deviations from the Schwarzschild values are moderate. The $\pi$ channel stays essentially at $+1$ over the whole range, and along this channel both the impact parameter and the periapsis remain nearly constant. The higher channels drift monotonically away from their Schwarzschild limits, and the drift grows with the channel number. At the largest couplings explored the $5\pi$ channel comes close to zero, and the $7\pi$ channel even changes sign.

For the self-dual black hole the two quantum parameters play very different roles. The minimum-area parameter $\ahat$ (Fig.~\ref{fig:cosxiqSDa}) enters the metric functions only at order $\ahat^2u^4$, and correspondingly its effect on $\cos\xi$ is mild. Over the whole range explored the strong-channel values move by only a few percent, and the $\pi$ channel, a near-parabolic orbit whose periapsis lies far outside the strong-field region, is flat to better than one part in $10^{8}$. The polymeric parameter $\cP$ (Fig.~\ref{fig:cosxiqSDb}) is the dominant deformation. At fixed ADM mass it moves the whole strong-field region outward, with $r_+=2M/(1-\cP)^2$, and it controls the mismatch $fg\neq1$ in \eqref{eq:fgSD}. The strong-channel values of $\cos\xi$ then drift monotonically away from their Schwarzschild limits as $\cP$ grows, the drift becoming a substantial fraction of the full range of $\cos\xi$ at the largest values explored, while the $\pi$ channel stays close to $+1$ throughout. This ordering is consistent with the structure of the transport coefficients discussed in Sec.~\ref{sec:spineq}. The self-dual coefficients \eqref{eq:spinSD} receive corrections already at order $\cP u$, linear in the quantum parameter, while those of the covariant metric are corrected only at order $\alpha^2u^2$.

\subsection{Dependence on the particle velocity}
\label{sec:vdep}

\begin{figure}[htpb]
	\centering
	\begin{subfigure}{0.48\textwidth}
		\centering
		\includegraphics[width=1.0\textwidth]{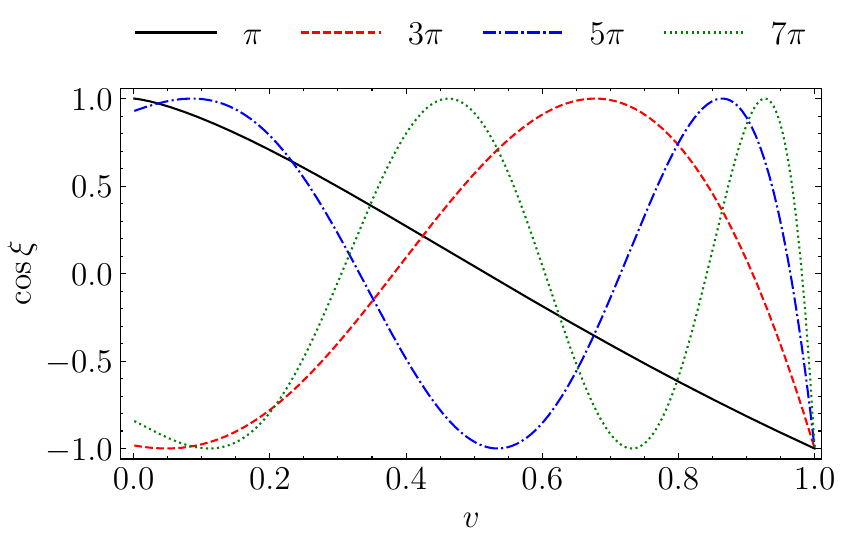}
		\caption{covariant, $\alpha=2$}
		\label{fig:cosxiv-a}
	\end{subfigure}\hfill
	\begin{subfigure}{0.48\textwidth}
		\centering
		\includegraphics[width=1.0\textwidth]{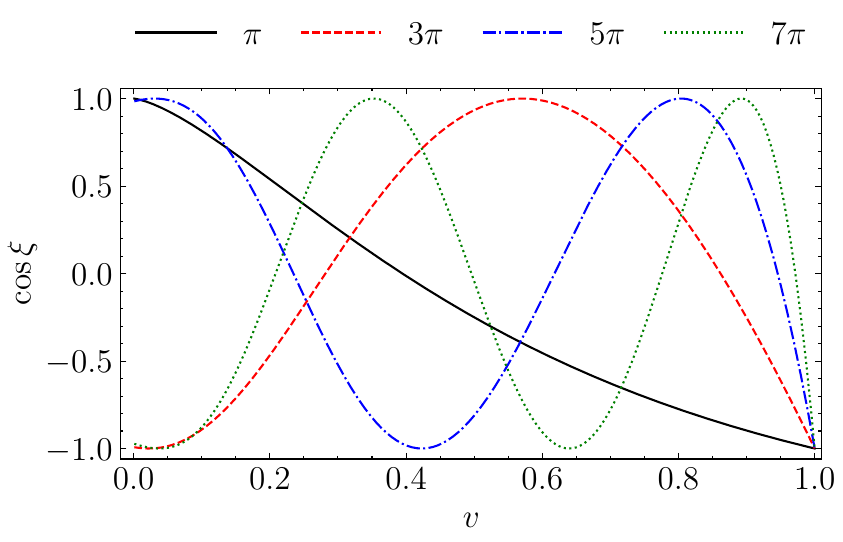}
		\caption{self-dual, $(\ahat,\cP)=(1,1/4)$}
		\label{fig:cosxiv-b}
	\end{subfigure}
	\caption{$\cos\xi$ as a function of the particle speed $v$, for the same deflection angles and the same line styles as in Fig.~\ref{fig:cosxiq}. Fig.~\ref{fig:cosxiv-a}: the covariant metric at $\alpha=2$; Fig.~\ref{fig:cosxiv-b}: the self-dual black hole at $(\ahat,\cP)=(1,1/4)$. All curves approach the universal spin flip $\cos\xi=-1$ as $v\to1$, and the $n$-th channel has exactly $n$ zeros in $0<v<1$.}
	\label{fig:cosxiv}
\end{figure}

Fig.~\ref{fig:cosxiv} shows the velocity dependence for the two spacetimes. Three features stand out. 

First, as $v\to1$ all curves approach $\cos\xi=-1$ for both spacetimes and all four channels, to better than one part in $10^{3}$ at the largest speeds computed. The universal spin flip of massless particles in backward scattering~\cite{Pang:2024tco} is therefore confirmed numerically for the loop quantum black holes, and the quantum corrections are invisible to particles moving at the speed of light. At high $v$ the orbits in the $3\pi$, $5\pi$ and $7\pi$ channels approach the photon sphere, and the impact parameter and the periapsis tend to the critical impact parameter and the photon-sphere radius of the respective spacetime.

Second, in the non-relativistic limit the curves converge to parameter-dependent values, listed in table~\ref{tab:vlimit}. In Schwarzschild spacetime the small-$v$ limit depends only on the deflection angle, $\cos\xi=(-1)^{n+1}$. The quantum corrections break this universal pattern. The size of the effect is controlled by the quantum parameters. At the moderate values of table~\ref{tab:vlimit} the deviations from the Schwarzschild values are at the percent level for the self-dual black hole, and larger for the covariant metric in the higher channels. For the self-dual black hole they grow rapidly with $\cP$ (cf.\ Fig.~\ref{fig:cosxiqSDb}) and become a substantial fraction of the Schwarzschild values themselves. In particular, for the dominant $\pi$ channel of backward scattering, slow particles have $\cos\xi$ close to $1$ in all three spacetimes, so the spin cancellation behind the missing glory spot does not operate and the glory spot survives~\cite{Pang:2024tco}. The deviations from unity, although small in this channel, show that the interference pattern in the backward direction is quantitatively modified by the quantum geometry.

\begin{table}[htpb]
	\centering
	\begin{tabular}{c|c|c|c}
		\hline
		$\Delta\varphi$ & Schwarzschild & covariant, $\alpha=2$ & self-dual, $(\ahat,\cP)=(1,1/4)$ \\
		\hline
		$\pi$ & $+1$ & $0.9998$ & $0.9997$ \\
		$3\pi$ & $-1$ & $-0.983$ & $-0.9927$ \\
		$5\pi$ & $+1$ & $0.929$ & $0.9842$ \\
		$7\pi$ & $-1$ & $-0.843$ & $-0.9731$ \\
		\hline
	\end{tabular}
	\caption{The small-$v$ limit of $\cos\xi$ for the four backward channels, compared with the Schwarzschild values $\cos\xi=(-1)^{n+1}$.}
	\label{tab:vlimit}
\end{table}

Third, $\cos\xi$ oscillates as a function of $v$, and the number of zeros in the interval $0<v<1$ equals the channel number, namely one zero for $\Delta\varphi=\pi$, two for $3\pi$, three for $5\pi$ and four for $7\pi$, for both spacetimes. Between consecutive zeros $\cos\xi$ reaches $\pm1$ within numerical accuracy, so the spin sweeps through an increasing number of full rotations as the channel number grows, before the universal flip takes over at $v\to1$. The positions of the zeros are metric-dependent and can in principle serve as fingerprints of the underlying geometry.

\subsection{The approach to the ultra-relativistic limit}
\label{sec:urslope}

The universal value $\cos\xi=-1$ at $v=1$ leaves open the question of how fast the flip is approached as $v\to1$. Since the transport equations \eqref{eq:spingen} are linear, their solution scales with the initial data, and it is convenient to write $(\tilde S^r,\bar S^\varphi)=E_0\,(s_r,s_\varphi)$, so that $(s_r,s_\varphi)$ solves the same equations with the velocity-independent initial conditions $s_r(0)=0$, $s_\varphi(0)=1$. In terms of these variables the definition \eqref{eq:cosxi} becomes
\begin{IEEEeqnarray}{rCl}
	\cos\xi &=& \frac{\operatorname{sgn}\left[s_\varphi(2n\pi)\right]}{\sqrt{1+\left(1-v^2\right)\left[s_r(2n\pi)/s_\varphi(2n\pi)\right]^2}} \quad . \label{eq:cosxisgn}
\end{IEEEeqnarray}
At $v=1$ the universal spin flip is the statement that $s_\varphi(2n\pi)<0$, and Eq.~\eqref{eq:cosxisgn} reduces to $\cos\xi=-1$ for any static, spherically symmetric spacetime. Expanding for $v\to1^{-}$ then gives
\begin{IEEEeqnarray}{rCl}
	\cos\xi &=& -1+\left(1-v\right)\left[\frac{s_r(2n\pi)}{s_\varphi(2n\pi)}\right]^{2}_{v=1}+\mathcal O\!\left((1-v)^{2}\right) \quad , \label{eq:cosxiexp}
\end{IEEEeqnarray}
and therefore
\begin{IEEEeqnarray}{rCl}
	\left.\frac{\dd(\cos\xi)}{\dd v}\right|_{v=1} &=& -\left[\frac{s_r(2n\pi)}{s_\varphi(2n\pi)}\right]^{2}_{v=1}\;\leq\;0 \quad . \label{eq:slopev1}
\end{IEEEeqnarray}
The slope is finite, non-positive, and determined entirely by the $v=1$ orbit: it is minus the squared ratio of the radial to the azimuthal spin component accumulated along the orbit, a ratio that drops out of $\cos\xi$ itself at $v=1$ because the radial component is suppressed by the boost factor $E_0^{-1}$. The value $\left.\cos\xi\right|_{v=1}=-1$ is thus universal, but the rate at which it is approached is not. Moreover, the slope is a strong-field quantity in every backward channel. At $v=1$ the orbits of all four channels whirl close to the photon sphere of the respective spacetime, whereas in the non-relativistic limit the $\pi$ channel is a near-parabolic orbit far outside the strong-field region. One should therefore expect visible quantum imprints on the slope in all four channels, including the $\pi$ channel whose small-$v$ value is essentially unaffected.

Each point below is computed by solving the boundary value problem of Sec.~\ref{sec:bvp} directly at $v=1$, where the orbit equation and the supplementary condition remain regular, continuing from the Schwarzschild-connected branch as before, and performing a single linear solve for the transported spin with the normalized initial data $(s_r,s_\varphi)(0)=(0,1)$. As an independent check, for a sample of points of both spacetimes and of Schwarzschild spacetime we recomputed the slope from $\cos\xi$ evaluated at $v=1-10^{-3}$ and $v=1-10^{-4}$, extrapolating linearly in $1-v$; the two determinations agree in all cases to better than one part in $10^{5}$, including for the highest channel.

\begin{figure}[htbp]
	\centering
	\includegraphics[width=0.62\textwidth]{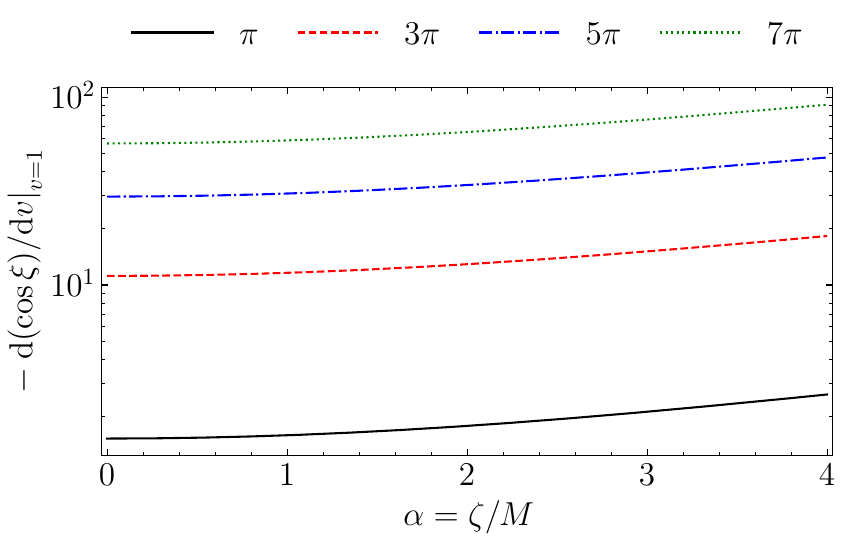}
	\caption{Minus the slope of the spin precession at $v=1$, $-\left.\dd(\cos\xi)/\dd v\right|_{v=1}$, as a function of the quantum parameter $\alpha=\zeta/M$ of the covariant metric, for the same deflection angles and the same line styles as in Fig.~\ref{fig:cosxiq}. Although $\cos\xi(1)=-1$ is universal, the rate at which it is approached depends visibly on $\alpha$ in all four channels. The Schwarzschild values at $\alpha=0$ are $1.52$, $11.2$, $29.6$ and $56.7$ respectively.}
	\label{fig:dcosxidv}
\end{figure}

\begin{figure}[htbp]
	\centering
    \begin{subfigure}{0.48\textwidth}
        \includegraphics[width=1.0\textwidth]{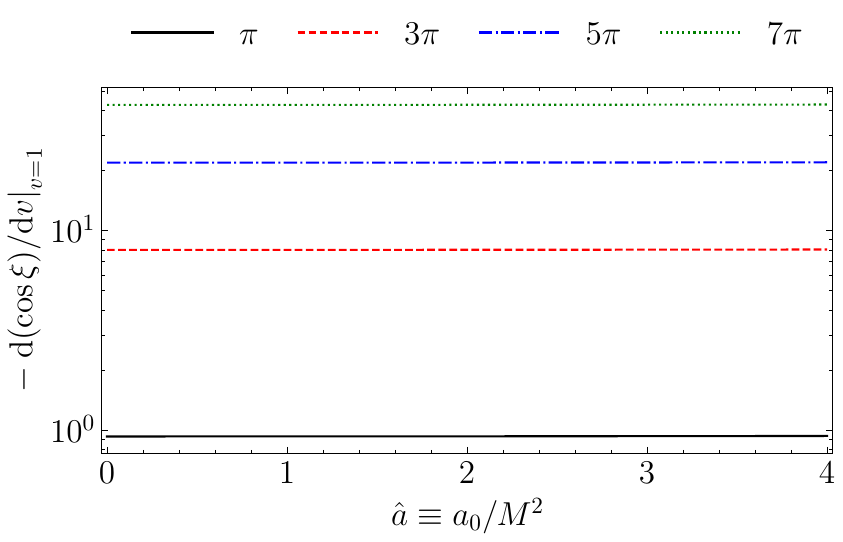}
        \caption{$\cP=1/4$}
        \label{fig:dcosxidvSDa}
    \end{subfigure}\hfill
    \begin{subfigure}{0.48\textwidth}
        \includegraphics[width=1.0\textwidth]{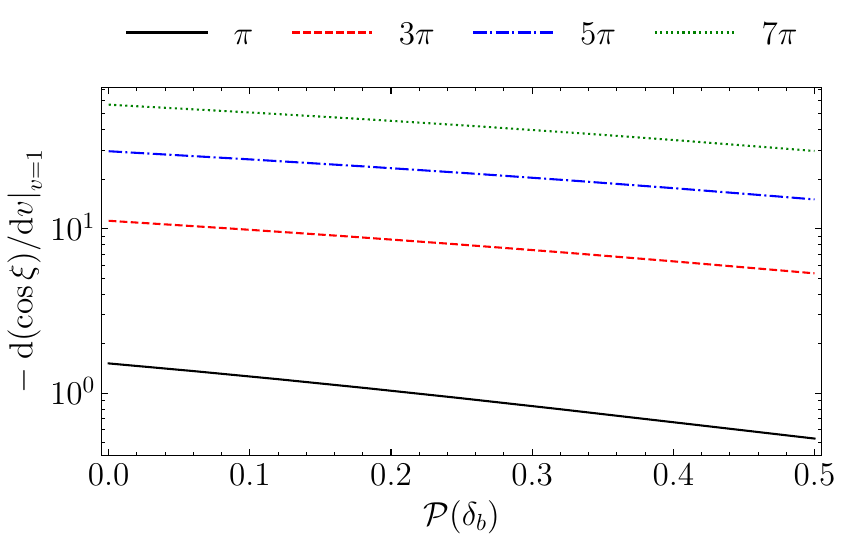}
        \caption{$\ahat=1$}
        \label{fig:dcosxidvSDb}
    \end{subfigure}
	\caption{Minus the slope of the spin precession at $v=1$ for the self-dual black hole, for the same deflection angles and the same line styles as in Fig.~\ref{fig:cosxiq}. Fig.~\ref{fig:dcosxidvSDa}: the dependence on the minimum-area parameter $\ahat=a_0/M^2$ at fixed $\cP=1/4$; Fig.~\ref{fig:dcosxidvSDb} the dependence on the polymeric parameter $\cP$ at fixed $\ahat=1$. The minimum-area parameter changes the slope by less than a percent, while the polymeric parameter reduces it by up to a factor of three.}
	\label{fig:dcosxidvSD}
\end{figure}

Figs.~\ref{fig:dcosxidv} and~\ref{fig:dcosxidvSD} show the negative of the slope \eqref{eq:slopev1} for the two spacetimes, over the same parameter ranges as in Sec.~\ref{sec:alphadep}. In Schwarzschild spacetime the slopes are $1.52$, $11.2$, $29.6$ and $56.7$ for the four channels, the rapid growth with the channel number reflecting the increasingly long whirl near the photon sphere. For the covariant metric (Fig.~\ref{fig:dcosxidv}) the magnitude of the slope grows monotonically with $\alpha$ in every channel.
For the self-dual black hole (Fig.~\ref{fig:dcosxidvSD}) the two quantum parameters again play very different roles. The minimum-area parameter $\ahat$ (Fig.~\ref{fig:dcosxidvSDa}) changes the slope by less than a percent over the whole range, consistent with its entering the metric functions only at order $\ahat^2u^4$. The polymeric parameter $\cP$ (Fig.~\ref{fig:dcosxidvSDb}) is the dominant deformation and drives the magnitude of the slope \emph{down}.
The opposite signs of the two effects fit the structure of the metrics. As we have discussed in Sec.~\ref{sec:spineq}, the deformation of the covariant metric deepens the strong-field region at fixed ADM mass, while for the self-dual black hole the polymeric parameter shifts the whole strong-field region outward. 

\section{Discussion}
\label{sec:discussion}

In this paper we have formulated gravitational lensing with a prescribed deflection angle as a two-point boundary value problem, and shown that pseudo-spectral methods solve it accurately at a modest computational cost. The impact parameter appears as a nonlinear eigenvalue determined together with the orbit, and the spin transport equations reduce to a single linear system on the same collocation grid. Each point in figures~\ref{fig:cosxiq}, \ref{fig:cosxiqSD} and~\ref{fig:cosxiv} requires only a small dense algebraic system with $53$ to $101$ grid points, and the results have been validated against independent shooting integrations. Although we focused on backward scattering, where the boundary interval is fixed by the deflection angle, nothing in the framework is specific to the two spacetimes studied. One only needs to supply the functions $f$, $g$ and $h$ of the metric~\eqref{eq:metric}. The self-dual black hole illustrates this generality. It shares with the covariant metric the mismatch between $g_{tt}$ and $-1/g_{rr}$ and in addition has an areal radius different from the radial coordinate, yet it is handled by the same machinery once the three functions are supplied.

Applied to the loop quantum black holes, the method reveals a clear picture of spin precession in backward scattering. Quantum corrections are absent for ultra-relativistic particles, whose spin is always flipped, but they survive and can be large for slowly moving massive particles. The velocity dependence displays a regular oscillation pattern with exactly $n$ zeros for the $n$-th backward channel. Furthermore, although the value $\cos\xi=-1$ at $v=1$ is universal, the rate at which it is approached is not: the slope $\dd(\cos\xi)/\dd v$ at $v=1$ is a photon-orbit quantity that depends visibly on the quantum parameters in all four backward channels of both spacetimes. For the self-dual black hole the polymeric parameter $\cP$ is the dominant deformation. At fixed ADM mass it pushes the strong-field region outward and drives the strong-channel values of $\cos\xi$ far from their Schwarzschild limits, while the minimum-area parameter $\ahat$ has only a mild effect. Since the spin precession controls the cancellation between contributions from perpendicular orbits, these results imply that the backward glory scattering of slowly moving spinning particles is sensitive to the quantum geometry, in contrast to the massless case where the glory spot vanishes universally. We remark that for astrophysical black holes the quantum parameters are expected to be tiny, as they are tied to the Planck scale~\cite{Zhang:2024khj,Modesto:2009ve,Huang:2025ggs}. Our scans up to $\alpha=4$, $\ahat=4$ and $\cP=1/2$ amplify the signal so that its structure becomes visible, while the small-parameter end of every curve is anchored at the Schwarzschild values.

Several directions are left for future work. It is natural to extend the framework to rotating quantum black holes, where the spin-orbit coupling makes the precession genuinely three-dimensional, to compute the glory cross sections with the spin precession taken into account, and to apply the same boundary value formulation to bound or nearly bound orbits, for instance the precession along eccentric orbits or the spinning-particle corrections à la Mathisson-Papapetrou.

\acknowledgments
XP is supported by the National Natural Science Foundation of China (NSFC) with Grant No. 12505082, and by the Doctoral Initiation Grant 24KE051 and Basic Research Grant 25kx010 from China West Normal University. X.G. is supported by Yancheng Institute of Technology (xjr2024030). 

\appendix 
\section{Pseudo-spectral methods and Chebyshev polynomials}
\label{sec:pseudospectral}
In this appendix we give a brief introduction to the pseudo-spectral (collocation) method and to the Chebyshev polynomials underlying it, focusing on the ingredients used in the main text, namely the representation of functions and derivatives on a Chebyshev-Gauss-Lobatto grid, the implementation of boundary conditions, and the treatment of nonlinear problems with a free parameter. For comprehensive presentations we refer to~\cite{Boyd:1989cfs,Trefethen:2000spectral,Grandclement:2007sb}, and for applications in gravitational contexts close to ours to~\cite{Guo:2016hiw,Andrade:2017jmt,Jansen:2017oag}. SubSec.~\ref{sec:collocation} collects the general machinery, and subSec.~\ref{sec:workedexample} works through a fully explicit example: the Schwarzschild orbit of the $5\pi$ backward channel, solved with $N=9$.

\subsection{The Chebyshev collocation method}
\label{sec:collocation}

\paragraph{Spectral expansions and interpolation.} The starting point of any spectral method is the representation of the unknown function as a finite expansion in smooth basis functions. In contrast to finite difference methods, which approximate derivatives locally on a small stencil and whose error decreases algebraically with the resolution, spectral methods are global. The value of the function at every grid point contributes to its derivative at every other point. When the solution is smooth on the interval of interest, the error decreases faster than any power of the inverse resolution, typically geometrically as $\ee^{-\mu N}$ for some constant $\mu>0$, a property known as spectral (or exponential) convergence~\cite{Boyd:1989cfs,Grandclement:2007sb}. The orbit and transport equations of this paper have rational coefficients and smooth solutions, and are therefore ideally suited to such methods.

On a finite interval a convenient basis is provided by the Chebyshev polynomials $T_n(x)$, defined on $x\in[-1,1]$ by $T_n(\cos\theta)=\cos(n\theta)$. Instead of the expansion coefficients it is usually more practical to work with the values of the function at a set of collocation points, the Chebyshev-Gauss-Lobatto points
\begin{IEEEeqnarray}{rCl}
	x_i &=& \cos\frac{i\pi}{N} \quad , \qquad i=0,1,\dots,N \quad , \label{eq:glpts}
\end{IEEEeqnarray} 
which are the extrema of $T_N$ together with the endpoints of the interval. Their clustering near the endpoints (the spacing is of order $N^{-2}$ there, versus $N^{-1}$ in the interior) suppresses the Runge phenomenon that spoils polynomial interpolation on equally spaced grids, and is one of the reasons for the good behaviour of the method~\cite{Trefethen:2000spectral}. The values at the grid points and the Chebyshev coefficients are related by the discrete Chebyshev transform, so that the two representations can be converted into each other as needed.

\paragraph{Cardinal functions and the differentiation matrix.} Associated with the collocation points are the cardinal (Lagrange) functions $C_j(x)$, polynomials of degree $N$ satisfying $C_j(x_i)=\delta_{ij}$; for the Chebyshev-Gauss-Lobatto points they can be written explicitly as~\cite{Jansen:2017oag}
\begin{IEEEeqnarray}{rCl}
	C_j(x)&=&\frac{2}{N p_j}\sum_{m=0}^N \frac{1}{p_m}T_{m}(x_{j})T_{m}(x)  \quad,\qquad p_0=p_N=2,~~p_j=1~~\text{otherwise} \quad . \label{eq:cardinal}
\end{IEEEeqnarray} 
Any function is then approximated by its interpolant, and its derivatives by the derivatives of the interpolant,
\begin{IEEEeqnarray}{rCl}
	f(x) &\simeq& \sum_{j=0}^N f(x_j)\,C_j(x) \quad , \qquad f'(x_i)\simeq\sum_{j=0}^N C_j'(x_i)\,f(x_j) \quad . \label{eq:interp}
\end{IEEEeqnarray} 
The second relation defines the differentiation matrix $D_{ij}=C_j'(x_i)$, the discrete analogue of the operator $\dd/\dd x$: differentiation of a function represented by its grid values is replaced by a single matrix multiplication.

The same matrix can be derived from the coefficient side. Expanding
\begin{IEEEeqnarray}{rCl}
	f(x) &=& \sum_{n=0}^N c_n T_n(x) \quad, \label{eq:fxcnTncontinuum}
\end{IEEEeqnarray} 
and evaluating at the collocation points gives $f(x_i)=\sum_{n=0}^N T_{in}c_n$ with $T_{in}=T_n(x_i)$, hence $c_n=\sum_{i=0}^N (T^{-1})_{ni}f(x_i)$. Differentiating the expansion and evaluating at the grid points,
\begin{IEEEeqnarray}{rCl}
	f'(x_i) &=& \sum_{n=0}^N c_n T_n'(x_i) \;=\; \sum_{j=0}^N\left[\sum_{n=0}^NT'_{in}(T^{-1})_{nj}\right] f(x_j) \quad,  \label{eq:fpxicnTnxicollo}
\end{IEEEeqnarray} 
so that
\begin{IEEEeqnarray}{rCl}
	D_{ij} &=& \sum_{n=0}^NT'_{in}(T^{-1})_{nj} \quad, \label{eq:DfromT}
\end{IEEEeqnarray} 
which indeed coincides with $C_j'(x_i)$. Its entries can be given in closed form~\cite{Trefethen:2000spectral}:
\begin{IEEEeqnarray}{rCl}
	D_{ij} &=& \frac{c_i}{c_j}\frac{(-1)^{i+j}}{x_i-x_j} \quad (i\neq j) \quad , \qquad D_{ii}=-\frac{x_i}{2\left(1-x_i^2\right)} \quad (1\leq i\leq N-1) \quad , \nonumber\\
	D_{00} &=& \frac{2N^2+1}{6} \;=\; -D_{NN} \quad , \label{eq:Dentries}
\end{IEEEeqnarray} 
with $c_0=c_N=2$ and $c_i=1$ otherwise. Higher order derivatives are obtained from matrix powers, $D^{(n)}=D^n$, and a general interval $[a,b]$ is covered by the linear map $x\mapsto\frac12\left[(b-a)x+(a+b)\right]$, with the chain rule supplying the corresponding Jacobian factors. The matrix $D$ is dense, but the resolutions needed in practice are modest (tens of points in the applications of this paper), so that the resulting linear algebra is inexpensive.

\paragraph{Boundary conditions.} A linear differential equation $Ly=f$ is thereby reduced to a linear algebraic system $D_L\,\bm y=\bm f$ for the vector $\bm y$ of values at the grid points. Boundary conditions are imposed by row replacement as follows.
\begin{enumerate}
	\item replace the rows of $D_L$ that correspond to the boundary points by the corresponding rows of the identity matrix (or by the collocated form of a more general boundary operator);
	\item replace the same entries of $\bm f$ by the prescribed boundary values.
\end{enumerate}
No reorganization of the basis is needed, and the same prescription applies unchanged to each Newton iterate when the equation is nonlinear, or to the initial conditions of the first order transport system of Sec.~\ref{sec:imppseudo}.

\paragraph{Nonlinear equations and free parameters.} For a nonlinear differential equation the collocation produces a nonlinear algebraic system $F(\bm U)=0$ for the values $\bm U$ at the grid points, which is solved by Newton's method. Starting from an initial guess $\bm U^{(0)}$, one iterates $\bm U^{(k+1)}=\bm U^{(k)}-J^{-1}F\big(\bm U^{(k)}\big)$, with the Jacobian $J=\partial F/\partial\bm U$ computed analytically from the differentiated equation. Convergence is typically quadratic once the iterate is close enough to a solution. If the continuum problem contains a free parameter (in the main text the squared normalized impact parameter $\eta^2$, which plays the role of a nonlinear eigenvalue), the parameter is appended to the vector of unknowns, and one additional scalar condition, here the first integral evaluated at the boundary, is added to the system, so that the discrete problem remains square. When a good initial guess is not available, parameter continuation, that is solving a sequence of nearby problems and chaining each converged solution as the next initial guess, provides a robust way to reach the target point and to follow a selected solution branch, as done in Sec.~\ref{sec:imppseudo}.

\subsection{A worked example: the Schwarzschild orbit}
\label{sec:workedexample}

We now work through one orbit of the main text explicitly: a massive particle with speed $v=1/2$, backward scattered by Schwarzschild spacetime in the channel $2\varphi_0=6\pi$, i.e.\ $\Delta\varphi=5\pi$ with $n=3$. The boundary value problem \eqref{eq:orbit}, \eqref{eq:orbitbc}, \eqref{eq:etacondition} specializes to
\begin{IEEEeqnarray}{rCl}
	u''+u &=& \frac{3}{4\,\eta^2}+3u^2 \quad , \qquad 0<\varphi<6\pi \quad , \nonumber\\
	u(0) &=& u(6\pi) \;=\; 0 \quad , \qquad \eta^2\,\big[u'(0)\big]^2 \;=\; \frac14 \quad , \label{eq:wkdbvp}
\end{IEEEeqnarray}
with the squared normalized impact parameter $\eta^2$ to be determined together with the orbit. The solution is a near-capture orbit: the particle arrives from infinity, whirls about two turns just outside the unstable circular orbit with the same energy ($r_c=2\sqrt3\,M$ for $E_0=2/\sqrt3$), and returns to infinity.

For Schwarzschild spacetime the boundary value problem is in fact exactly solvable in terms of elliptic integrals, which provides the exact reference for the comparisons below. The first integral is the cubic
\begin{IEEEeqnarray}{rCl}
	\big[u'(\varphi)\big]^2 &=& \frac{1}{\eta^2}\left[v^2+2u\left(1-v^2\right)\right]-u^2\left(1-2u\right) \;=\; 2\left(u-u_t\right)\left(u-u_2\right)\left(u-u_3\right) \quad , \label{eq:wkdcubic}
\end{IEEEeqnarray}
with three real roots ordered $u_3<0<u_t<u_2$. The smallest positive root is the turning point, and it fixes the eigenvalue through
\begin{IEEEeqnarray}{rCl}
	\eta^2 &=& \frac{v^2+2u_t\left(1-v^2\right)}{u_t^2\left(1-2u_t\right)} \quad , \label{eq:wkdeta2ut}
\end{IEEEeqnarray}
while the remaining two roots follow from the sum and product relations $u_2+u_3=\frac12-u_t$ and $u_2u_3=-v^2/\left(2\eta^2u_t\right)$. Integrating the cubic gives the orbit in closed form,
\begin{IEEEeqnarray}{rCl}
	u(\varphi) &=& u_3+\left(u_t-u_3\right)\operatorname{sn}^2\!\left(\omega\,\varphi+F(\chi_0,k)\,,\,k\right) \quad , \qquad \omega=\sqrt{\frac{u_2-u_3}{2}} \quad , \label{eq:wkdexact}
\end{IEEEeqnarray}
with $\operatorname{sn}$ the Jacobi elliptic function, $F$ the incomplete elliptic integral of the first kind, modulus $k^2=\left(u_t-u_3\right)/\left(u_2-u_3\right)$, and $\sin^2\chi_0=-u_3/\left(u_t-u_3\right)$. The boundary condition at $\varphi=6\pi$ then reads
\begin{IEEEeqnarray}{rCl}
	\sqrt{\frac{2}{u_2-u_3}}\,\left[K(k)-F(\chi_0,k)\right] &=& 3\pi \quad , \label{eq:wkdret}
\end{IEEEeqnarray}
with $K(k)=F(\pi/2,k)$ the complete elliptic integral. This is an exact transcendental equation for the turning point: it cannot be solved in closed form, but its numerical solution determines $u_t$, and with it $\eta^2$ through \eqref{eq:wkdeta2ut}, to any desired accuracy. For the present parameters this gives $u_t=0.28833400$ and $\eta^2=19.39235174$, with $k^2=0.99814$ close to the circular-orbit separatrix $k^2=1$, as expected for a near-capture orbit. For the loop quantum metrics of Sec.~\ref{sec:metrics} the first integral is a sextic polynomial for the covariant metric and a rational function for the self-dual black hole, rather than a cubic, and no such closed form exists.

We use $N=9$: the grid is $x_j=\cos(j\pi/9)$, mapped to the orbit interval by $\varphi_j=3\pi(1-x_j)$. The first differentiation matrix, computed from the closed form \eqref{eq:Dentries}, is small enough to be written down explicitly (its entries are exact algebraic numbers, e.g.\ $D^{(1)}_{00}=(2N^2+1)/6=163/6$; we quote three decimals):
{\setlength{\arraycolsep}{2.4pt}\scriptsize
\begin{equation}
	D^{(1)}=\begin{pmatrix}
		27.170 & -33.160 & 8.549 & -4.000 & 2.420 & -1.704 & 1.333 & -1.132 & 1.031 & -0.500 \\
		8.291 & -4.017 & -5.759 & 2.274 & -1.305 & 0.898 & -0.695 & 0.586 & -0.532 & 0.258 \\
		-2.137 & 5.759 & -0.927 & -3.759 & 1.688 & -1.064 & 0.790 & -0.653 & 0.586 & -0.283 \\
		1.000 & -2.274 & 3.759 & -0.333 & -3.064 & 1.484 & -1.000 & 0.790 & -0.695 & 0.333 \\
		-0.605 & 1.305 & -1.688 & 3.064 & -0.090 & -2.879 & 1.484 & -1.064 & 0.898 & -0.426 \\
		0.426 & -0.898 & 1.064 & -1.484 & 2.879 & 0.090 & -3.064 & 1.688 & -1.305 & 0.605 \\
		-0.333 & 0.695 & -0.790 & 1.000 & -1.484 & 3.064 & 0.333 & -3.759 & 2.274 & -1.000 \\
		0.283 & -0.586 & 0.653 & -0.790 & 1.064 & -1.688 & 3.759 & 0.927 & -5.759 & 2.137 \\
		-0.258 & 0.532 & -0.586 & 0.695 & -0.898 & 1.305 & -2.274 & 5.759 & 4.017 & -8.291 \\
		0.500 & -1.031 & 1.132 & -1.333 & 1.704 & -2.420 & 4.000 & -8.549 & 33.160 & -27.170
	\end{pmatrix} \quad , \label{eq:wkdD1}
\end{equation}
}%
\noindent and $D^{(2)}=(D^{(1)})^2$; the map to the orbit interval supplies the factors $\dd/\dd\varphi=-(3\pi)^{-1}D^{(1)}$ and $\dd^2/\dd\varphi^2=(9\pi^2)^{-1}D^{(2)}$.

The orbit is symmetric about the turning point, $u(\varphi)=u(6\pi-\varphi)$, so only the first half of the grid is independent, as in Sec.~\ref{sec:imppseudo}: the unknowns are $(u_0,u_1,u_2,u_3,u_4)$ and $\eta^2$, the full vector of values is reconstructed by even reflection, $(u_0,u_1,\dots,u_4,u_4,\dots,u_1,u_0)$, and the second boundary condition $u(6\pi)=0$ is automatic. The discrete system is six equations for six unknowns,
\begin{IEEEeqnarray}{rCl}
	u_0 &=& 0 \quad , \nonumber\\
	\frac{1}{9\pi^2}\sum_{j=0}^{4}\left[D^{(2)}_{ij}+D^{(2)}_{i,9-j}\right]u_j+u_i-3u_i^2-\frac{3}{4\eta^2} &=& 0 \quad , \qquad i=1,\dots,4 \quad , \nonumber\\
	\eta^2\left[\sum_{j=0}^{4}\left(D^{(1)}_{0j}+D^{(1)}_{0,9-j}\right)u_j\right]^2 &=& \frac{9\pi^2}{4} \quad , \label{eq:wkddiscrete}
\end{IEEEeqnarray}
namely the boundary condition at $\varphi=0$, the collocated orbit equation at the four interior points of the half grid, and the supplementary condition fixing $\eta^2$, in which the first row of $D^{(1)}$ supplies $u'(0)$. With the numerical values of \eqref{eq:wkdD1} the last equation reads, for instance,
\begin{IEEEeqnarray}{rCl}
	\eta^2\left(26.670\,u_0-32.129\,u_1+7.417\,u_2-2.667\,u_3+0.716\,u_4\right)^2 &=& 88.826 \quad . \label{eq:wkdsupp}
\end{IEEEeqnarray}

\begin{figure}[htpb]
	\centering
	\includegraphics[width=0.72\textwidth]{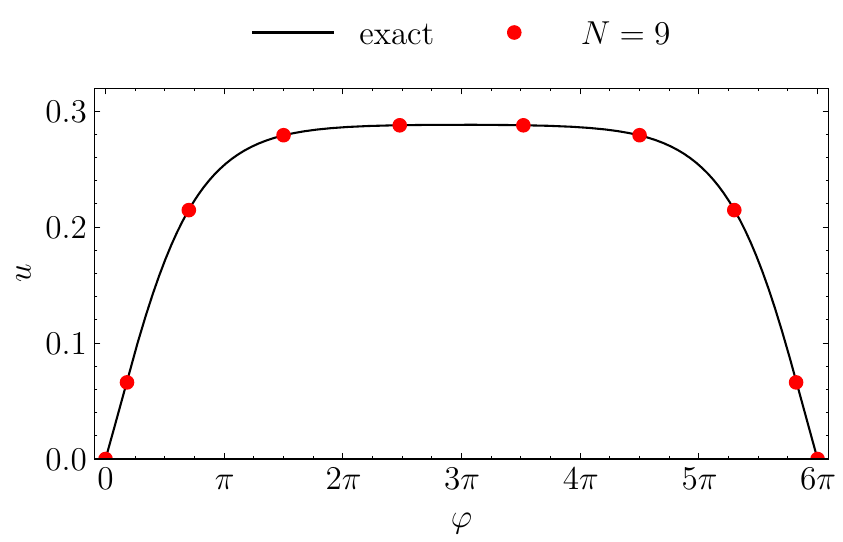}
	\caption{The orbit of the worked example of appendix~\ref{sec:workedexample}: the exact elliptic-function solution \eqref{eq:wkdexact} (solid curve) and the coarse computation with $N=9$ (points). The total swept angle $2\varphi_0=6\pi$ corresponds to the deflection angle $\Delta\varphi=2\varphi_0-\pi=5\pi$.}
	\label{fig:wkdorbit}
\end{figure}

Newton's method with the analytically computed Jacobian is started from the initial guess $u_j^{(0)}=\frac13\tanh\varphi_j$, a smooth profile plateauing at the photon-sphere value. The initial value of $\eta^2$ is taken from the circular geodesic with the same energy, $\eta^2_{\rm circ}=(1-v^2)/\left[u_c\left(1-3u_c\right)\right]$ with $u_c=\sqrt3/6$, i.e.\ $\eta^2=9+6\sqrt3\simeq19.3923$. The iteration converges quadratically: the residual norm over the successive iterates is $2.8$, $4.5\times10^{-1}$, $8.5\times10^{-2}$, $7.0\times10^{-3}$, $5.8\times10^{-5}$, $4.2\times10^{-9}$, $3.1\times10^{-16}$, each step roughly squaring the error of the previous one, and the round-off level is reached after six iterations. The converged solution is
\begin{IEEEeqnarray}{rCl}
	(u_0,u_1,u_2,u_3,u_4) &=& (0\;,\;0.06610\;,\;0.21470\;,\;0.27931\;,\;0.28786) \quad , \nonumber\\
	\eta^2 &=& 19.385985 \quad , \qquad \eta \;=\; 4.40295 \quad , \label{eq:wkdsol}
\end{IEEEeqnarray}
with $u'(0)=0.113560$. The turning point $\varphi=3\pi$ is not a grid point for odd $N$; the turning value follows from the interpolant \eqref{eq:interp}, $u_t=u(3\pi)=\sum_j u_j\,C_j(0)=0.29089$, corresponding to the periapsis $r_{\rm min}=3.44\,M$. The solution is compared with the exact orbit \eqref{eq:wkdexact} in Fig.~\ref{fig:wkdorbit}: the coarse grid values already follow the exact orbit closely.

How accurate is the coarse answer? The exact reference values of \eqref{eq:wkdret} are reproduced to twenty digits by the same spectral computation with $N=73$, whose differentiation matrix can no longer usefully be written down; the agreement also cross-checks the exact solution. The $N=9$ computation already reproduces the nonlinear eigenvalue to $0.03\%$ and the turning value to about a percent. The refinement between the two resolutions is shown in table~\ref{tab:wkdconv}: after the expected coarse-grid wiggles the error decreases geometrically with $N$, the spectral convergence of subSec.~\ref{sec:collocation}, until it reaches the level of the working precision.

\begin{table}[htpb]
	\centering
	\begin{tabular}{c|c|c|c|c}
		\hline
		$N$ & $\eta^2$ & $|\delta\eta^2|/\eta^2$ & $u_t$ & $|\delta u_t|/u_t$ \\
		\hline
		$9$ & $19.385985$ & $3.3\times10^{-4}$ & $0.29089388$ & $8.9\times10^{-3}$ \\
		$11$ & $19.502118$ & $5.7\times10^{-3}$ & $0.28823290$ & $3.5\times10^{-4}$ \\
		$13$ & $19.450110$ & $3.0\times10^{-3}$ & $0.28842096$ & $3.0\times10^{-4}$ \\
		$15$ & $19.402577$ & $5.3\times10^{-4}$ & $0.28840856$ & $2.6\times10^{-4}$ \\
		$21$ & $19.392080$ & $1.4\times10^{-5}$ & $0.28833399$ & $2.9\times10^{-8}$ \\
		$31$ & $19.392351$ & $2.2\times10^{-8}$ & $0.28833399$ & $4.5\times10^{-9}$ \\
		$45$ & $19.39235174$ & $1.7\times10^{-12}$ & $0.2883339949$ & $2.2\times10^{-13}$ \\
		$73$ & $19.39235174$ & $\sim10^{-21}$ & $0.2883339949$ & $\sim10^{-22}$ \\
		\hline
		exact \eqref{eq:wkdexact} & $19.3923517437$ &  & $0.2883339949$ &  \\
		\hline
	\end{tabular}
	\caption{Refinement of the worked example with the resolution $N$. $\eta^2$ is the nonlinear eigenvalue of the boundary value problem \eqref{eq:wkdbvp} and $u_t$ the turning value $u(3\pi)$, evaluated from the interpolant \eqref{eq:interp}. Errors are measured relative to the exact elliptic-integral solution \eqref{eq:wkdexact}, shown in the last row; the $N=73$ row is limited by the working precision of the spectral computation.}
	\label{tab:wkdconv}
\end{table}

The orbits of the main text are computed in exactly the same way; only the coefficients of the equation change, from the constants of the Binet-type Eq.~\eqref{eq:wkdbvp} to rational functions of the metric functions $f$, $g$ and $h$, and the spin transport equations of Sec.~\ref{sec:spineq} are then solved on the same grid with a single linear solve, as described in Sec.~\ref{sec:imppseudo}.

\providecommand{\href}[2]{#2}\begingroup\raggedright\endgroup

\end{document}